\documentclass[12pt]{article}
\usepackage{graphicx}
\usepackage{graphics}
\title{Integral and series representations of the digamma and polygamma functions}
\author{Mark W. Coffey\\
Department of Physics\\
Colorado School of Mines\\
Golden, CO  80401\\
(Received $\mbox{~~~~~~~~~~~~~~~~~~~~~~~~~~~~~~~2010}$)}
\date{August 10, 2010}
\begin{document}
\maketitle
%\vspace{.25cm}
\baselineskip=25 pt
\begin{abstract}

We obtain a variety of series and integral representations of the digamma function
$\psi(a)$.  These in turn provide representations of the evaluations $\psi(p/q)$ at
rational argument and for the polygamma function $\psi^{(j)}$.  The approach is through a limit definition of the zeroth Stieltjes constant $\gamma_0(a)=-\psi(a)$.  
Several other results are obtained, including product representations for $\exp[\gamma_0(a)]$
and for the Gamma function $\Gamma(a)$.  In addition, we present series representations in terms of trigonometric integrals Ci and Si for $\psi(a)$ and the Euler constant $\gamma=-\psi(1)$.
%then how to do for $\gamma_j(a)$, $j \geq 1$ more generally?

\end{abstract}
 
%\vspace{.20cm}
\medskip
\baselineskip=15pt
\centerline{\bf Key words and phrases}
\medskip 

\noindent

Gamma function, digamma function, polygamma function, Euler constant, series representation, integral representation, Hurwitz zeta function, cosine integral, sine integral 

%\vspace{.25cm}
\vfill
\centerline{\bf 2010 AMS codes} 
33B15, 33C20, 11Y60

\baselineskip=25pt
\pagebreak
\medskip
\centerline{\bf Introduction and statement of results}
\medskip

We recall the defining Laurent expansion of the Stieltjes constants $\gamma_k(a)$\cite{coffeyjmaa,coffey2009,coffeyjnt,coffeystdiffs,stieltjes,wilton2}
$$\zeta(s,a)={1 \over {s-1}}+\sum_{n=0}^\infty {{(-1)^n} \over {n!}}\gamma_n(a)(s-1)^n,
\eqno(1.1)$$
where $\zeta(s,a)$ is the Hurwitz zeta function.  Notationally, we let $\zeta(s)=
\zeta(s,1)$ be the Riemann zeta function \cite{edwards,ivic,riemann,titch}, $\Gamma$ the 
Gamma function, $\psi=\Gamma'/\Gamma$ be the digamma function (e.g., \cite{nbs}) with $\gamma=-\psi(1)$ the Euler constant, $\psi^{(k)}$ be the polygamma functions \cite{nbs}, and $_pF_q$ be the generalized hypergeometric function \cite{andrews}. 
%may well use later:
%%In addition, $P_1(x)=B_1(x-[x])=x-[x]-1/2$ denotes the first periodized Bernoulli %%polynomial, with $\{x\}=x-[x]$ the fractional part of $x$.  
%Being periodic, $P_1$ has the
%Fourier series \cite{nbs} (p. 805)
%$$P_1(x)=-\sum_{j=1}^\infty {{\sin 2\pi j x} \over {\pi j}}.  \eqno(1.1)$$

The Stieltjes constants may be expressed through the limit relation \cite{berndt}
$$\gamma_n(a)={{(-1)^n} \over {n!}} \lim_{N\to \infty}\left[\sum_{k=0}^N {{\ln^n (k+a)}
\over {k+a}}-{{\ln^{n+1} (N+a)} \over {n+1}}\right], ~~~~~~n \geq 0.  \eqno(1.2)$$
Here, $a \notin \{0,-1,-2,\ldots\}$.  For an asymptotic expression for these constants,
even valid for moderate values of $n$, \cite{knessl2} (Section 2) may be consulted.

In this paper, we obtain various representations of the digamma function via the
connection $\gamma_0(a)=-\psi(a)$ \cite{wilton2}.  These in turn lead to many special cases, including the values $\psi(p/q)$ for rational argument, and further imply representations of
the polygamma functions.  We obtain product representations for $\exp[\gamma_0(a)]$ and
$\Gamma(a)$.  
%see de Doelder, Grosjean (too?) refs ...for polygamma & Cl_2 evaluations
We present series representations for $\ln \Gamma(a)$, $\psi(a)$, and $\gamma$ using the trigonometric integrals Si and Ci.  In addition, we provide several summations over 
parameterized values of Ci.

The following is an example of representations that we develop.
\newline{\bf Proposition 1}.  Let Re $a>0$.  Then we have 
$$\gamma_0(a)=-\psi(a)={1 \over 2}\int_0^\infty {1 \over {(t+a)(t+a+1)}} ~_3F_2(1,2,2;
3,t+a+2;1)dt-\ln a \eqno(1.3)$$
$$=-\ln a-\sum_{k=1}^\infty {1 \over {k+1}}\sum_{\ell=0}^k (-1)^\ell{k \choose \ell} 
\ln(\ell+a) \eqno(1.4)$$
$$=-\ln a-\int_0^1 u^{a-1}\left({1 \over {u-1}}-{1 \over {\ln u}}\right)du \eqno(1.5)$$
$$=-\ln a-\int_0^1\int_0^1 {{(xy)^{a-1} (1-x)} \over {(1-xy)\ln(xy)}}dx dy \eqno(1.6a)$$
$$=-\ln a-\int \int_T {{(1-Y)^{a-1}(1-X)} \over {XY \ln(1-Y)}}dXdY \eqno(1.6b)$$
$$=-\ln a-\int_0^\infty \sum_{k=2}^\infty {1 \over k^2}{1 \over {{t+k+a-1} \choose k}}dt.  \eqno(1.6c)$$
In (1.6b), $T$ is the triangle with vertices at $(1,0)$, $(0,1)$, and $(1,1)$.

Thus from (1.4) we have
\newline{\bf Corollary 1}.  We have for Re $a>0$
$$e^{\gamma_0(a)}={1 \over a}\prod_{k=1}^\infty \prod_{\ell=0}^k \left[(\ell+a)^{(-1)^{
\ell+1}{k \choose \ell}} \right]^{1 \over {k+1}}.  \eqno(1.7)$$
Proposition 1 and this Corollary subsume expressions for $\gamma$ given in \cite{sondow}.

In addition, from Proposition 1 follow representations for the polygamma functions, and these
include
\newline{\bf Corollary 2}.  We have for Re $a>0$
$$\psi^{(j)}(a)=(-1)^{j-1}{{(j-1)!} \over a^j}+(j-1)!\sum_{k=1}^\infty {1 \over {k+1}}\sum_{\ell=0}^k (-1)^\ell{k \choose \ell} {{(-1)^{j-1}} \over {(\ell+a)^j}}
\eqno(1.8)$$
$$=(-1)^{j-1}{{(j-1)!} \over a^j}+\int_0^1 u^{a-1}\left({1 \over {u-1}}-{1 \over {\ln u}}\right)\ln^j u ~du \eqno(1.9)$$
$$=(-1)^{j-1}{{(j-1)!} \over a^j}+\int_0^1\int_0^1 {{(xy)^{a-1} (1-x)} \over {(1-xy)}}
\ln^{j-1}(xy) ~dx dy \eqno(1.10a)$$
$$=(-1)^{j-1}{{(j-1)!} \over a^j}+\int \int_T {{(1-Y)^{a-1}(1-X)} \over {XY}}\ln^{j-1}(1-Y) dXdY. \eqno(1.10b)$$

Hence we obtain representations at positive integer arguments for harmonic numbers $H_n 
\equiv \sum_{k=1}^n 1/k$ and generalized harmonic numbers $H_n^{(r)} \equiv \sum_{k=1}^n 1/k^r$.  For we have $H_n = \psi(n+1)-\psi(1) = \psi(n+1)+\gamma$ and
$$H_n^{(r)}={{(-1)^{r-1}} \over {(r-1)!}}\left[\psi^{(r-1)}(n+1)-\psi^{(r-1)}
(1) \right ], \eqno(1.11)$$
where $\psi^{(r-1)}(1)=(-1)^r(r-1)!\zeta(r)$. 

Further, if we integrate $-\int_1^a \gamma_0(b)db=\ln \Gamma(a)$ we have
\newline{\bf Corollary 3}.  We have for Re $a>0$
$$\ln \Gamma(a)=a(\ln a-1)+1+\sum_{k=1}^\infty {1 \over {k+1}}\sum_{\ell=0}^k (-1)^\ell{k \choose \ell} [(\ell+a)\ln(\ell+a)- (\ell+1)\ln(\ell+1)] \eqno(1.12)$$
$$=a(\ln a-1)+1+\int_0^1 (u^{a-1}-1)\left({1 \over {u-1}}-{1 \over {\ln u}}\right){{du } \over {\ln u}} \eqno(1.13a)$$
$$=a(\ln a-1)-{1 \over 2}\ln a+{1 \over 2}\ln(2\pi)+\int_0^1 u^{a-1}\left({1 \over {u-1}}-{1 \over {\ln u}}+{1 \over 2}\right){{du } \over {\ln u}} \eqno(1.13b)$$
$$=a(\ln a-1)+1+\int_0^1\int_0^1 {{[(xy)^{a-1}-1] (1-x)} \over {(1-xy)\ln^2(xy)}}dx dy \eqno(1.14a)$$
$$=a(\ln a-1)+1+\int \int_T {{[1-(1-Y)^{a-1}](1-X)} \over {XY \ln^2(1-Y)}}dXdY. \eqno(1.14b)$$
By inspection, we see that the right sides of (1.12)-(1.14) properly vanish at $a=1$.
The term $a(\ln a-1)$ on the right sides there is the leading asymptotic form of $\ln
\Gamma(a)$ when $a \to \infty$.  In writing (1.13b), we have used (e.g., \cite{sri}, p. 16)
$$\int_0^1\left({1 \over {u-1}}-{1 \over {\ln u}}+{1 \over 2}\right){{du} \over {\ln u}}
=1-{1 \over 2} \ln(2\pi), ~~~~\int_0^1 {{u^{a-1}-1} \over {\ln u}}du=\ln a, ~~\mbox{Re}
~a>0, \eqno(1.15)$$
thereby recovering Binet's first expression for $\ln \Gamma(a)$.

Then we have from (1.12)
\newline{\bf Corollary 4}.  We have for Re $a>0$
$$\Gamma(a)=e^{1-a}a^a\prod_{k=1}^\infty \prod_{\ell=0}^k \left[\left[{{(\ell+a)^{\ell+a}}
\over {(\ell+1)^{\ell+1}}}\right]^{(-1)^\ell{k \choose \ell}} \right]^{1 \over {k+1}}.  \eqno(1.16)$$

A multitude of summation formulas for the digamma function is known (e.g., \cite{hansen},
Section 55, \cite{coffey05jcam}).  Our integral representations in particular well provide a basis for further developing such summations.  We simply give an example here.
In light of the asymptotic relation as $a \to \infty$,
$$\psi(a)=\ln a -{1 \over {2a}}-{1 \over {12 a^2}}+O\left({1 \over a^4}\right), \eqno(1.17)$$
we may consider the following.
\newline{\bf Corollary 5}.  For Re $\alpha>0$ and Re $(\alpha+\beta)>0$, we have (a)
$$\sum_{n=1}^\infty\left[\psi(\alpha n+\beta)-\ln(\alpha n+\beta)+{1 \over {2(\alpha n+\beta)}}+{1 \over {12(\alpha n+\beta)^2}}\right]$$
$$=\int_0^1 {v^{\beta/\alpha} \over {1-v}}\left[{1 \over {\alpha(v^{1/\alpha}-1)}}-{1 \over {\ln v}}+{1 \over {2\alpha}}-{1 \over {12\alpha^2}}\ln v\right]dv,  \eqno(1.18)$$
and (b) for $\alpha \gg 1$
$$\sum_{n=1}^\infty\left[\psi(\alpha n+\beta)-\ln(\alpha n+\beta)+{1 \over {2(\alpha n+\beta)}}\right] \sim -\sum_{k=1}^\infty {1 \over \alpha^{2k-1/2}}{B_{2k} \over {2k}}
\zeta\left(2k,1+{\beta \over \alpha}\right),$$
where $B_j$ is the $j$th Bernoulli number.

Then we may determine the asymptotic dependence to all orders of a certain second
moment of the Riemann xi function.  For this, we put $\xi(s)=(s-1)\pi^{-s/2}\Gamma(1+s/2)
\zeta(s)$, and $\Xi(t) \equiv \xi(1/2+it)$.  We have
\newline{\bf Corollary 6}.  We have as $\alpha \to \infty$
$$\int_0^\infty \left|\Xi\left({t \over 2}\right)\Gamma\left({{it-1} \over 4}\right)
\right|^2 {{\cos[(t/2)\ln \alpha]} \over {t^2+1}}dt$$
$$\sim {\pi^{3/2} \over {2\sqrt{\alpha}}}[\ln \alpha+\ln(2\pi)-\gamma]+\pi^{3/2}
\sum_{k=1}^\infty {1 \over \alpha^{2k-1/2}}{B_{2k} \over {2k}}
\zeta\left(2k\right).$$
The moment integral here, going back to Ramanujan, is of interest from many points of
view \cite{berndtd}.

Let 
$$\mbox{Ci}(z) \equiv -\int_z^\infty {{\cos t} \over t}dt, ~~~~~~~~\mbox{Si}(z) \equiv \int_0^z {{\sin t} \over t}dt. \eqno(1.19)$$
Then we have
\newline{\bf Proposition 2}. We have
$$\gamma={1 \over 2}+ \mbox{Ci}(\pi)+{1 \over {2\pi}}\left\{\sum_{j=1}^\infty {1 \over {j+1}}
[\pi-2\mbox{Si}(\pi j)]+\sum_{j=2}^\infty {1 \over {j-1}}[\pi-2\mbox{Si}(\pi j)]\right\}.
\eqno(1.20)$$
In this expression, Ci$(\pi) \simeq 0.07366079$ and therefore the sum terms provide 
small corrections.  %discuss more on convergence of the sums and asymptotic forms in
%Remarks in Section 2.  To compare + contrast this Prop. 1 with the repr(s) in my notes
%wrt Nathan McKenzie's notes.

{\bf Proposition 3}. We have for Re $a>0$
$$\psi(a)=\ln a-{1 \over {2a}}+\sum_{j=1}^\infty \left[2\cos(2\pi ja)\mbox{Ci}(2\pi ja)
-\sin(2\pi ja)[\pi-2\mbox{Si}(2\pi ja)]\right].  \eqno(1.21)$$

Then from $\psi(n+1)=\psi(n)+1/n$ we have
\newline{\bf Corollary 7}.  We have
$$H_n=\ln n+\gamma+{1 \over {2n}}+2\sum_{j=1}^\infty \mbox{Ci}(2\pi n j).$$
Alternatively, this result follows from the Euler-Maclaurin summation expression
$$H_n=\ln n+\gamma+{1 \over {2n}}+\int_n^\infty {{P_1(x)} \over x^2}dx
=\ln n+\gamma+{1 \over {2n}}+{1 \over n}\int_1^\infty {{P_1(nv)} \over v^2}dv, $$
where $P_1$ is the polynomial given in (2.29).  %see p. 60 of my notes for proof of latter
Similarly, for $r>1$ we have
$$H_n^{(r)}=\zeta(r)+{1 \over {2n^r}}-{n^{1-r} \over {r-1}}+r\int_n^\infty {{P_1(x)} \over x^{r+1}}dx
=\zeta(r)+{1 \over {2n^r}}-{n^{1-r} \over {r-1}}+{r \over n^r}\int_1^\infty {{P_1(nv)} \over v^{r+1}}dv. $$
%see p. 61 of my notes of 7/30/10 for proof of the latter relation.

Let $B_n(x)$ be the Bernoulli polynomial of degree $n$, $B_n=B_n(0)$ the $n$th Bernoulli
number, and $B_n(1/2)=(2^{1-n}-1)B_n$ \cite{nbs} (p. 805).  Then we have the following.
{\bf Proposition 4}. Let $0 < \beta \leq 2 \pi$.  Then we have (a)
$$\sum_{n=1}^\infty {{\mbox{Ci}(\beta n)} \over n^2}=(\gamma+\ln \beta)\zeta(2)-\zeta'(2)
-{\pi \over 2}\beta +{\beta^2 \over 8}, \eqno(1.22a)$$
$$\sum_{n=1}^\infty (-1)^n {{\mbox{Ci}(\beta n)} \over n^2}={1 \over 2}(\ln2-\gamma-\ln \beta)\zeta(2) +{1\over 2}\zeta'(2)+{\beta^2 \over 8}, \eqno(1.22b)$$
(b)
$$\sum_{n=1}^\infty {{\mbox{Ci}(\beta n)} \over n^4}=(\gamma+\ln \beta)\zeta(4)-\zeta'(4)
+{\beta^2 \over {12}}\left(-{\beta^2 \over {16}}+{\pi \over 3}\beta-{\pi^2 \over 2}\right),
\eqno(1.22c)$$
$$\sum_{n=1}^\infty (-1)^n{{\mbox{Ci}(\beta n)} \over n^4}=(\ln 2-7\gamma-7\ln \beta) {{\zeta(4)} \over 8}+{7 \over 8}\zeta'(4)
-{\beta^4 \over {192}}+{{\zeta(2)} \over 8}\beta^2,  \eqno(1.22d)$$
(c) for integers $k \geq 1$
$$\sum_{n=1}^\infty {{\mbox{Ci}(\beta n)} \over n^{2k}}=(\gamma+\ln \beta)\zeta(2k)-\zeta'(2k)
+{{(-1)^{k-1}} \over {2(2k)!}}(2\pi)^{2k}\int_0^1 {1 \over v}\left[B_{2k}\left({{\beta v} \over {2\pi}}\right)-B_{2k}\right]dv, \eqno(1.23a)$$
%for $k \geq 1$
$$\sum_{n=1}^\infty (-1)^n {{\mbox{Ci}(\beta n)} \over n^{2k}}=(\gamma+\ln \beta)(2^{1-2k}-1)\zeta(2k)+(1-2^{1-2k})\zeta'(2k)+2^{1-2k}(\ln 2)\zeta(2k)$$
$$+{{(-1)^{k-1}} \over {2(2k)!}}(2\pi)^{2k}\int_0^1 {1 \over v}\left[B_{2k}\left({{\beta v+\pi} \over {2\pi}}\right)-B_{2k}\left({1 \over 2}\right)\right]dv, \eqno(1.23b)$$
(d) for Re $a>1$,
$$\sum_{n=1}^\infty {{\mbox{Ci}(\beta n)} \over n^a}=(\gamma+\ln \beta)\zeta(a)-\zeta'(a)
+\int_0^1\left\{{{(2\pi)^a} \over {4\Gamma(a)}}\sec\left({{\pi a} \over 2}\right)\left[\zeta\left(1-a,1-{{\beta v} \over {2\pi}}\right)+\zeta\left(1-a,{{\beta v} \over {2\pi}}\right)\right]\right.$$
$$\left. -\zeta(a)\right\} {{dv} \over v}, \eqno(1.24a)$$
$$\sum_{n=1}^\infty (-1)^n {{\mbox{Ci}(\beta n)} \over n^a}=(\gamma+\ln \beta)(2^{1-a}-1)\zeta(a)+(1-2^{1-a})\zeta'(a)+2^{1-a}\zeta(a)\ln 2$$
$$+\int_0^1\left\{{{(2\pi)^a} \over {4\Gamma(a)}}\sec\left({{\pi a} \over 2}\right)\left[\zeta\left(1-a, {{\pi-\beta v} \over {2\pi}}\right)+\zeta\left(1-a,{{\pi+\beta v} \over {2\pi}}\right)\right] +(1-2^{1-a})\zeta(a)\right\}{{dv} \over v}, \eqno(1.24b)$$
(e) for Re $a>1$,
$$\sum_{n=1}^\infty {1 \over {(2n+1)^a}} \mbox{Ci}(\beta n)=(\gamma+\ln \beta) [(1-2^{-a})\zeta(a)-1]+S(a)$$
$$+\int_0^1 \left\{ {{(2\pi)^a} \over {4\Gamma(a)}}\csc(\pi a)\left[-\sin\left({{\beta v+\pi a}\over 2}\right)\left[\zeta\left(1-a,{{\beta v+2\pi} \over {4\pi}}\right)-\zeta\left(1-a,{{\beta v} \over {4\pi}}\right)\right] \right.\right.$$
$$\left. \left. +\sin\left({{\beta v-\pi a}
\over 2}\right)\left[\zeta\left(1-a,{{2\pi-\beta v} \over {4\pi}}\right)-\zeta\left(1-a,1-
{{\beta v} \over {4\pi}}\right)\right]\right]
+(2^{-a}-1)\zeta(a) \right\}{{dv} \over v},  \eqno(1.25a)$$
$$\sum_{n=1}^\infty {{(-1)^n} \over {(2n+1)^a}} \mbox{Ci}(\beta n)=(\gamma+\ln \beta) [4^{-a}(\zeta(a,1/4)-\zeta(a,3/4))-1]+T(a)$$
$$+\int_0^1 \left\{ {{(2\pi)^a} \over {4\Gamma(a)}}\csc(\pi a)\left[\cos\left({{\beta v+\pi a}
\over 2}\right)\left[\zeta\left(1-a,{{\beta v+\pi} \over {4\pi}}\right)-\zeta\left(1-a,{{\beta v+3\pi} \over {4\pi}}\right)\right] \right.\right.$$
$$\left. \left. +\cos\left({{\beta v-\pi a}
\over 2}\right)\left[\zeta\left(1-a,{{\pi-\beta v} \over {4\pi}}\right)-\zeta\left(1-a,{{3 \pi-\beta v} \over {4\pi}}\right)\right]\right]
+4^{-a}[\zeta(a,3/4)-\zeta(a,1/4)] \right\}{{dv} \over v},  \eqno(1.25b)$$
where,
$$S(a) \equiv \sum_{n=2}^\infty {{\ln n} \over {(2n+1)^a}}, ~~~~~~T(a) \equiv \sum_{n=2}^\infty {{(-1)^n \ln n} \over {(2n+1)^a}},  \eqno(1.26)$$
and (f) for $|z|\leq 1$,
$$\sum_{n=1}^\infty {z^n \over {n(n+1)}}\mbox{Ci}(\beta n)=(\gamma+\ln \beta)\left[1-\ln(1-z)
+{1 \over z}\ln(1-z)\right]+\sum_{n=1}^\infty {{z^n \ln n} \over {n(n+1)}}$$
$$+{1 \over z}\int_0^1 \left[(z-1)\ln(1-z)]-{1 \over 2}(z-\cos \beta v)\ln(1-2z\cos \beta v
+z^2)-\sin \beta v \tan^{-1}\left({{z \sin \beta v} \over {1-z\cos \beta }}\right)\right] {{dv} \over v}. \eqno(1.27)$$

\medskip
%\pagebreak
\centerline{\bf Proof of Propositions}

We let $\psi'$ be the trigamma function and $(b)_n=\Gamma(b+n)/\Gamma(b)$ be the
Pochhammer symbol.  

{\it Proposition 1}.  Preliminary relations are contained in
\newline{\bf Lemma 1}.  We have
$$\gamma_0(a)=\sum_{k=0}^\infty \int_0^\infty \left[{1 \over {(t+k+a)^2}}-{1 \over {(t+k+a+1)(t+k+a)}}
\right]dt-\ln a$$
$$=\sum_{k=0}^\infty \int_0^\infty {1 \over {(t+k+a)^2}}{{dt} \over {(t+k+a+1)}}-\ln a$$
$$=\int_0^\infty \left[\psi'\left(t+a\right)-{1 \over {t+a}}\right]dt-\ln a=-\psi(a).  \eqno(2.1)$$

{\it Proof}.  From relation (1.2) we have
$$\gamma_0(a)=\lim_{N \to \infty}\sum_{k=0}^N \left[{1 \over {k+a}}-\ln(N+a)\right]$$
$$=\lim_{N \to \infty}\sum_{k=0}^N \left[{1 \over {k+a}}-\ln\left({{k+a+1} \over {k+a}}
\right)\right]-\ln a$$
$$=\sum_{k=0}^\infty \int_0^\infty \left[{1 \over {(t+k+a)^2}}-{1 \over {(t+k+a+1)(t+k+a)}} \right]dt-\ln a. \eqno(2.2)$$
The rest of the Lemma follows easily upon noting (e.g., p. 259 of \cite{nbs})
$$\psi'(z)=\zeta(2,z)=\sum_{n=0}^\infty {1 \over {(n+z)^2}}.  \eqno(2.3)$$

We now write from the Lemma
$$\gamma_0(a)=\int_0^\infty {1 \over {(t+a)^2}}{1 \over {(t+a+1)}}\sum_{k=0}^\infty {{(t+a)^2
(t+a+1)} \over {(t+k+a)^2(t+k+a+1)}}dt-\ln a.  \eqno(2.4)$$
The integrand of (2.2) being absolutely convergent, the interchange of summation and
integration is justified.  In order to achieve hypergeometric form, we note the ratios
$${{(t+a)_k} \over {(t+a+1)_k}}={{t+a} \over {t+a+k}}, ~~~~~~~~ 
{{(t+a)_k} \over {(t+a+2)_k}}={{(t+a)(t+a+1)} \over {(t+a+k)(t+a+k+1)}}.  \eqno(2.5)$$
Upon using the series definition of the function $_3F_2$, we therefore obtain
$$\gamma_0(a)=\int_0^\infty {1 \over {(t+a)^2}}{1 \over {(t+a+1)}} ~_3F_2(1,t+a,t+a;t+a+1,
t+a+2;1)dt-\ln a$$
$$={1 \over 2}\int_0^\infty {1 \over {(t+a)^2}}{1 \over {(t+a+1)}} ~_3F_2(1,2,2;3,t+a+2;1)dt-\ln a.  \eqno(2.6)$$
Here, we have used the transformation \cite{sondow} (6), valid for Re $s>0$ and Re $v-t>0$,
$$_3F_2(1,s,t;s+1,v;1)={s \over {v-t}} ~_3F_2(1,v-t,v-s;v-t+1,v;1).  \eqno(2.7)$$
We have obtained (1.3).
%next use pp. 3 and 10 of my notes ...

Next we have
$$_3F_2(1,2,2;3,t+a+2;1)=\sum_{j=0}^\infty {{(2)_j^2} \over {(3)_j (t+a+2)_j}}
=2\sum_{j=0}^\infty {{(j+1)!} \over {(j+2)(t+a+2)_j}}$$
$$=2\sum_{j=1}^\infty {1 \over {j+1}}{{j!} \over {(t+a+2)_{j-1}}}
=2\sum_{j=1}^\infty {1 \over {j+1}}{{j!(t+a+j+1)} \over {(t+a+2)_j}}. \eqno(2.8)$$
The integral in (2.6) becomes
$${1 \over 2}\int_0^\infty {{~_3F_2(1,2,2;3,t+a+2;1)} \over {(t+a)^2}}{{dt} \over {(t+a+1)}}   =\int_0^\infty \sum_{j=1}^\infty {{j!} \over {(j+1)}} {{(t+a+j+1) dt} \over {(t+a)(t+a+1)
(t+a+2)_j}}. \eqno(2.9)$$
%$$=\int_0^\infty \sum_{j=1}^\infty {{j!} \over {j+1}}{{dt} \over {(t+a)(t+a+1)(t+a+2)\cdots
%(t+a+j)}} $$
Therefore, from (2.6) we find
$$\gamma_0(a)=-\ln a+\int_0^\infty \sum_{k=1}^\infty {{k!} \over {(k+1)(t+a)(t+a+1) \cdots
(t+k+a)}}dt.  \eqno(2.10)$$
We can carry out the integration by using the partial fraction decomposition \cite{gkp}
$${{N!} \over {x(x+1)\cdots (x+N)}} = \sum_{k=0}^N {N \choose k} {{(-1)^k} \over {(x+k)}}.
\eqno(2.11)$$
We then have (1.4).
%next to p. 8 and 9 notes.  

If we employ the Beta function integral
$$\int_0^1 u^{t+a-1}(1-u)^k du=B(t+a,k+1)={{k!} \over {(t+a)(t+a+1)_k}}, \eqno(2.12)$$
in (2.10) we find
$$\gamma_0(a)=-\ln a+\int_0^\infty \sum_{k=1}^\infty \int_0^1 {{u^{t+a-1}(1-u)^k} \over {k+1}}
dudt \eqno(2.13)$$
$$=-\ln a+\int_0^\infty \int_0^1 u^{t+a-1}\left({{\ln u} \over {u-1}}-1\right)du dt
. \eqno(2.14)$$
Performing the integral over $t$ gives (1.5).

We may rewrite (2.13) as
$$\gamma_0(a)=-\ln a+\int_0^\infty \int_0^1 \int_0^{1-u} {u^{t+a-1} \over {(1-u)}}{v \over
{(1-v)}}dvdudt$$
$$=-\ln a+\int_0^1 \int_0^{1-u} {u^{a-1} \over {(u-1)}} {v \over {(1-v)}}{{dvdu} \over {\ln u}}.  \eqno(2.15a)$$
We now put $u=xy$, $v=1-x$ and use the Jacobian $\partial(u,v)/\partial(x,y)=x$ to
obtain (1.6a).

Putting $x=X$ and $y=(1-Y)/X$, with Jacobian of transformation $\partial(X,Y)/\partial(x,y)
=-x$, in (1.6a) yields (1.6b).  Using $-1/\ln(1-Y)=\int_0^\infty (1-Y)^t dt$ in (1.6b) gives
$$\gamma_0(a)=-\psi(a)=-\ln a+\int \int_T \int_0^\infty {{(1-Y)^{t+a-1}} \over {XY}}(1-X) dtdXdY$$
$$=-\ln a+\int_0^\infty \int_0^1 \int_{1-Y}^1 {{(1-X)} \over {XY}}(1-Y)^{t+a-1} dXdYdt$$
$$=-\ln a-\int_0^\infty \int_0^1 {{(1-Y)^{t+a-1}} \over Y} [Y+\ln(1-Y)] dYdt \eqno(2.15b)$$
$$=-\ln a-\int_0^\infty \sum_{k=2}^\infty {1 \over k}\int_0^1 (1-Y)^{t+a-1}Y^{k-1}dYdt$$
$$=-\ln a-\int_0^\infty \sum_{k=2}^\infty {1 \over k}B(k,t+a)dt,$$
where we employed the Beta function integral of (2.12).  Then we have the expressions
$$\gamma_0(a)=-\ln a-\int_0^\infty \sum_{k=2}^\infty {1 \over k}{{\Gamma(k)\Gamma(t+a)} \over
{\Gamma(t+k+a)}}dt$$
$$=-\ln a-\int_0^\infty \sum_{k=2}^\infty {{\Gamma(k)} \over k}{{dt} \over {(t+a)_k}}, $$
that are equivalent to (1.6c).

{\it Corollary 5}.  From (1.5) we have
$$\psi(a)=\ln a-{1 \over {2a}}-{1 \over {12a^2}}+\int_0^1 u^{a-1}\left({1 \over {u-1}}-{1 \over {\ln u}}+{1 \over 2}-{1 \over {12}}\ln u\right)du. \eqno(2.16)$$
We then interchange summation and integration to find
$$\sum_{n=1}^\infty\left[\psi(\alpha n+\beta)-\ln(\alpha n+\beta)+{1 \over {2(\alpha n+\beta)}}+{1 \over {12(\alpha n+\beta)^2}}\right]$$
$$=\int_0^1 {u^{\alpha+\beta-1} \over {1-u^\alpha}}\left[{1 \over {u-1}}-{1 \over {\ln u}}
+{1 \over 2}-{1 \over {12}}\ln u\right]du. \eqno(2.17)$$
Making the change of variable $v=u^\alpha$ gives part (a) of the Corollary.
For part (b), we use the asymptotic relation for the digamma function as $z \to \infty$
$$\psi(z)=\ln z-{1 \over {2z}}-\sum_{k=1}^\infty {B_{2k} \over {2kz^{2k}}}.$$

{\it Corollary 6}.  This follows from the $\beta=0$ case of Corollary 5(b), together
with the reciprocity relation with $\alpha$, $\beta>0$ and $\alpha \beta=1$ \cite{berndtd}
$$\sqrt{\alpha}\left[{{\gamma-\ln(2\pi \alpha)} \over {2\alpha}}+\sum_{n=1}^\infty
\phi(n\alpha)\right]=\sqrt{\beta}\left[{{\gamma-\ln(2\pi \beta)} \over {2\beta}} +\sum_{n=1}^\infty \phi(n\beta)\right]$$
$$=-{1 \over \pi^{3/2}}\int_0^\infty \left|\Xi\left({t \over 2}\right)\Gamma\left({{it-1} \over 4}\right)\right|^2 {{\cos[(t/2)\ln \alpha]} \over {t^2+1}}dt,$$
where $\phi(x)=\psi(x)+1/2x-\ln x$.

An alternative approach is to expand the integrand of Corollary 5(a), using
$$v^{1/\alpha}-1=\sum_{j=1}^\infty {{\ln^j v} \over {j!\alpha^j}},$$
followed by the use of the integrals $\int_0^1 [\ln^j v/(1-v)]dv=(-1)^j j! \zeta(j+1)$.

We note that the reciprocity relation itself provides the complementary asymptotic
relation as $\alpha \to 0$.  For then $\beta=1/\alpha \to \infty$.

We also note that the leading term of the asymptotic relation in Corollary 6 is 
connected with the skew self-reciprocal inverse Fourier cosine transform
$${\cal F}_c^{-1}\left({{\ln \alpha} \over \sqrt{\alpha}}\right)=-{1 \over \sqrt{t}}
\left(\ln t+\gamma +{\pi \over 2}+2\ln 2\right).$$
This transform may be calculated by logarithmic differentiation with respect to $x$ of
the integral
$\sqrt{2 \over \pi}\int_0^\infty \alpha^x \cos(\alpha t)d\alpha$ with $-1<\mbox{Re} ~x
<0$, at $x=-1/2$.

{\it Remarks}.  If in (1.5) we put $u=\exp(-t)$, we have
$$\gamma_0(a)=-\ln a+\int_0^\infty e^{-at}\left({1 \over {1-e^{-t}}}-{1 \over t}\right)dt,
\eqno(2.18)$$
thereby recovering formula 8.361.8 of \cite{grad} (p. 943) for the digamma function.  This
formula is also recovered if we use an integral representation for $\ln$,
$\ln z=\int_0^\infty(e^{-t}-e^{-zt})(dt/t)$ for Re $z>0$, in (1.4).  For then we have
$$\gamma_0(a)=-\ln a+\sum_{k=1}^\infty {1 \over {k+1}}\int_0^\infty {e^{-at} \over t}(1-
e^{-t})^kdt$$
$$=-\ln a-\int_0^\infty e^{-at}\left({1 \over t}-{1 \over {1-e^{-t}}}\right)dt. \eqno(2.19)$$

If in (2.15b) we instead carry out the integration over $Y$, we recover the integral of (2.1).

Special values of $\gamma_0$ are $\gamma_0(1/2)=\gamma+2\ln 2$, and %p. 20 of Sri/Choi book
$$\gamma_0\left(n+{1 \over 2}\right)=\gamma+2\ln 2-2\sum_{k=0}^{n-1} {1 \over {2k+1}}.  \eqno(2.20)$$
Further, $\gamma_0(1/4)=\gamma+\pi/2+3\ln 2$, and $\gamma(3/4)=\gamma-\pi/2+3\ln 2$, giving for instance $\gamma_0(1/4)-\gamma_0(3/4)=\pi$.  More generally, for rational arguments we have the following (\cite{andrews}, p. 13 or \cite{sri}, p 19).  
\newline{\bf Theorem}.  
$$\psi\left({p \over q}\right)=-\gamma-{\pi \over 2}\cot{{\pi p} \over q}-\ln q+2\sum_{n=1}
^{[q/2]}{'} \cos{{2\pi np} \over q}\ln\left(2\sin{{\pi n} \over q}\right), \eqno(2.21)$$
where $0<p<q$; $\sum'$ means that when $q$ is even the term with index $n=q/2$ is divided by
$2$.  
%to mention we have gotten reprs for harmonic numbers--now done in the Intro section.
\newline{Therefore, we have found representations for all the values of (2.21).

The digamma and polygamma functions satisfy many properties including functional equations, duplication and multiplication formulas, and reflection formulas.  All such properties must 
be inherent in our various series and integral representations.
%later, to look into recovering various refl. formula, dupl. formula, 
%mult. formula etc. properties for the digamma function via these reprs ...
%+ all the many, many summation properties of \psi that are known--maybe pick out
%something that way to recover .....
As an illustration, we have
\newline{\bf Corollary 8} (multiplication formula).  For integers $m \geq 1$ we have
$$\gamma_0(ma)=-\ln m+{1 \over m}\sum_{k=0}^{m-1} \gamma_0\left(a+{k \over m}\right). \eqno(2.22)$$

{\it Proof}.  From (1.5) we have
$$\gamma_0(ma)=-\ln m-\ln a-\int_0^1 u^{ma-1}\left({1 \over {u-1}}-{1 \over {\ln u}}\right)du $$
$$=-\ln m-\ln a-\int_0^1 v^{a-1}\left[{1 \over {m(v^{1/m}-1)}}-{1 \over {\ln v}}\right]dv.
\eqno(2.23)$$
The Corollary then follows from the factorization
$${1 \over {v^{1/m}-1}}={{\sum_{k=0}^{m-1} v^{k/m}} \over {v-1}}.  \eqno(2.24)$$

%later look at \gamma_j(a)'s via generalized di/poly gamma functions ... TO DO ...

{\it Proposition 2}.  We have previously obtained the integral representation \cite{coffey2009} (2.86)
$$\gamma=\int_{-\infty}^\infty {{e^z \ln(1+e^{-z})} \over {z^2+\pi^2}}dz.  \eqno(2.25)$$
The idea of the proof is to suitably expand the logarithm of the integrand, and then to
perform termwise integration.  For this we write
$$\gamma=\left(\int_0^\infty+\int_{-\infty}^0\right) {{e^z \ln(1+e^{-z})} \over {z^2+\pi^2}}dz$$
$$=\sum_{j=1}^\infty {{(-1)^{j-1}} \over j} \int_0^\infty {e^{-(j-1)z} \over {z^2+\pi^2}}dz
+\int_0^\infty e^{-y}{{[y+\ln(1+e^{-y})]} \over {y^2+\pi^2}}dy$$
$$={1 \over 2}+\mbox{Ci}(\pi)+{1 \over {2\pi}}\sum_{j=2}^\infty {1 \over j}(\pi+2\mbox{Si}
[\pi(1-j)])
+\sum_{j=1}^\infty {{(-1)^{j-1}} \over j} \int_0^\infty {e^{-(j+1)y} \over {y^2+\pi^2}}dy$$
$$={1 \over 2}+\mbox{Ci}(\pi)+{1 \over {2\pi}}\sum_{j=2}^\infty {1 \over j}(\pi-2\mbox{Si}
[\pi(j-1)])
+{1 \over {2\pi}}\sum_{j=1}^\infty {1 \over j}(\pi-2\mbox{Si}[\pi(j+1)]).  \eqno(2.26)$$
Shifting the index in the summations gives the expression (1.20).

{\it Remarks}.  The asymptotic forms of Si and Ci are easily obtained to any order by
repeated integration by parts.  It is then easy to see that the summations in (1.20) have
summands that are $O(1/j^2)$, and additionally these leading terms have sign alternation
according to $(-1)^j$.

%then briefly compare with other result from my notes ...
For comparison purposes, we recall an earlier result \cite{cofunpub09}
$$1-\gamma={1 \over 2}+2\sum_{j=1}^\infty \mbox{Ci}(2\pi j), \eqno(2.27)$$ 
that readily shows how to develop $1-\gamma$ from $1/2$ with a series of corrections, and the leading terms in the corrections are easily written.

We recall that the constant (e.g., \cite{sri}, p. 345)
$$1-\gamma=\int_1^\infty {{\{t\}} \over t^2}dt,  \eqno(2.28)$$
where the fractional part $\{t\}=t-[t]$.  
We have that $P_1(x)=B_1(x-[x])=x-[x]-1/2$, the first periodized Bernoulli polynomial, has the standard Fourier series \cite{nbs} (p. 805),
$$P_1(x)=-\sum_{j=1}^\infty {{\sin(2\pi jx)} \over {\pi j}}.   \eqno(2.29)$$
Inserted into (2.22), we have
$$\int_1^\infty {{\{t\}} \over t^2}dt={1 \over 2}-{1 \over \pi}\sum_{j=1}^\infty
{1 \over j}\int_1^\infty \sin(2\pi j t) {{dt} \over t^2}$$
$$={1 \over 2}+2\sum_{j=1}^\infty \mbox{Ci}(2\pi j),  \eqno(2.30)$$
where we integrated by parts and made a simple change of variable.  

As a second proof of (2.27) we have the following.  We write
$$\sum_{j=1}^\infty \mbox{Ci}(2\pi j)=-\sum_{j=1}^\infty \int_{2\pi j}^\infty {{\cos t} \over
t} dt=-\sum_{j=1}^\infty \int_0^\infty {{\cos (v+2\pi j)} \over {v+2\pi j}} dv=
-\sum_{j=1}^\infty \int_0^\infty {{\cos v} \over {v+2\pi j}} dv$$
$$=-\sum_{j=1}^\infty \int_0^\infty \cos v dv\int_0^\infty e^{-(v+2\pi j)x}dx
=-\int_0^\infty \cos v \int_0^\infty {{e^{-xv} dx} \over {e^{2\pi x}-1}}$$
$$=-\int_0^\infty {{x dx} \over {(x^2+1)(e^{2\pi x}-1)}}={1 \over 2}\left({1 \over 2}-\gamma
\right).  \eqno(2.31)$$
Here the interchange of summation and integration is justfied by the absolute convergence
of the $x$-integration.
In the last step, we applied Hermite's expression for the digamma function (\cite{sri},
p. 91 or \cite{grad}, 8.361.3, p. 943).

Further representations for combinations of the Euler constant and zeta values may be 
obtained from the following.
\newline{\bf Lemma 2}.  We have for $k \geq 2$
$$I_k\equiv \int_1^\infty {{\{t\}^k} \over t^{k+1}}dt=1-\gamma-\sum_{j=2}^k {{[\zeta(j)-1]} 
\over j}.  \eqno(2.32)$$

We provide an operational proof, using the Dirac delta function $\delta$.  We have, 
integrating by parts,
$$I_k={1 \over k}\left[\int_1^\infty {1 \over t^k}\left({d \over {dt}}\{t\}^k\right)dt
-\left.{{\{t\}^k} \over t^k}\right|_1^\infty\right]$$
$$=\int_1^\infty {{\{t\}^{k-1}} \over t^k}dt-\int_1^\infty {{\{t\}^{k-1}} \over t^k}d[t]
+{1 \over k}$$
$$=I_{k-1}-\sum_{j=2}^\infty {1 \over j^k}+{1 \over k}.  \eqno(2.33)$$
Herein, we used $d[t]=\sum_j \delta(t-j)dt$.
%consider to give another proof--see 2009 notes ...

Of course $I_k \to 0$ as $k \to \infty$, and we have the simple bound $I_k \leq 1/k$.

We let $B_k(x)$ be the Bernoulli polynomials and $P_k(x)=B_k(x-[x])$ their periodized
form.  In applying Lemma 2 we may use the relation
$$\{x\}^n=(x-[x])^n={1 \over {n+1}}\sum_{k=0}^n {{n+1} \choose k}P_k(x), \eqno(2.34)$$
with $B_0(x)=1$, 
and the Fourier series for $n \geq 1$ \cite{nbs} (p. 805)
$$P_{2n-1}(x)=(-1)^n {{2(2n-1)!} \over {(2\pi)^{2n-1}}}\sum_{j=1}^\infty {{\sin(2\pi jx)}
\over j^{2n-1}}, ~~~~P_{2n}(x)=(-1)^{n-1} {{2(2n)!} \over {(2\pi)^{2n}}}\sum_{j=1}^\infty {{\cos(2\pi jx)} \over j^{2n}}.  \eqno(2.35)$$
Then we have
\newline{\bf Corollary 9}.  We have (a)
$$I_2={3 \over 2}-\gamma-{1 \over 2}\zeta(2)={1 \over 4}+\sum_{j=1}^\infty[2\mbox{Ci}(2\pi j)
+j\pi^2-1-2j\pi \mbox{Si}(2\pi j)], \eqno(2.36a)$$
and (b)
$$I_3={{11} \over 6}-\gamma-{1 \over 2}\zeta(2)-{1 \over 3}\zeta(3)={1 \over 4}+\sum_{j=1}^\infty\left[2\left(1-{2 \over 3}j^2\pi^2\right)\mbox{Ci}(2\pi j)-{4 \over 3}+j\pi^2-2j\pi \mbox{Si}(2\pi j)\right]. \eqno(2.36b)$$

{\it Proof}.  For part (a) we use the combination 
$$\{x\}^2=P_2(x)+P_1(x)+{1 \over 3}=\sum_{j=1}^\infty\left[{{\cos(2\pi jx)} \over {\pi^2 j^2}}
-{{\sin(2\pi jx)} \over {\pi j}}\right]+{1 \over 3}, \eqno(2.37)$$
and carry out the integrations.  For part (b) we again use the Fourier series (2.35), so
that 
$$\{x\}^3=P_3(x)+{3 \over 2}P_2(x)+P_1(x)+{1 \over 4}=\sum_{j=1}^\infty\left[{3 \over 2}{{\cos(2\pi jx)} \over {\pi^2 j^2}}+\left({3 \over {2\pi^2 j^2}}-1\right){{\sin(2\pi jx)}  \over {\pi j}}\right]+{1 \over 4}. \eqno(2.38)$$

%The summands of (2.36) and (2.37) are $O(1/j^2)$ for $j \to \infty$.

{\it Proposition 3}.  From (e.g., \cite{edwards}, p. 107)
$$\ln \Gamma(a)=\left(a-{1 \over 2}\right)\ln a -a+{1 \over 2}\ln (2\pi)-\int_0^\infty
{{P_1(t)} \over {t+a}}dt, \eqno(2.39)$$
we have
$$\psi(a)=\ln a-{1 \over {2a}}+\int_0^\infty {{P_1(t)} \over {(t+a)^2}}dt. \eqno(2.40)$$
Using the Fourier representation (2.29) and performing the integration gives the
Proposition.

{\it Remarks}.  When $a=1$ in (1.21) we recover (2.27), while from the case $a=1/2$ we find
the companion relation
$$\gamma+\ln 2=1-2\sum_{j=1}^\infty (-1)^j \mbox{Ci}(\pi j).  \eqno(2.41)$$
The sum on the right side of this equation may be determined similarly to how we found
(2.31).

Similar series representations in terms of Si and Ci may be obtained for $\ln \Gamma(a)$
and for the polygamma functions $\psi^{(j)}(a)$.  For instance, from (2.39) we have
$$\ln \Gamma(a)=\left(a-{1 \over 2}\right)\ln a -a+{1 \over 2}\ln (2\pi)$$
$$+{1 \over {2\pi}}\sum_{j=1}^\infty {1 \over j}\left[2\sin(2\pi ja)\mbox{Ci}(2\pi ja)
+\cos(2\pi ja)[\pi-2\mbox{Si}(2\pi ja)]\right].  \eqno(2.42)$$
%so implies such series reprs for the harmonic and gen. harmonic numbers, etc. too
From this equation at $a=1$ we may conclude that
$${1 \over {2\pi}}\sum_{j=1}^\infty {1 \over j}[\pi-2\mbox{Si}(2\pi j)]=1-{1 \over 2}\ln
(2\pi),  \eqno(2.43)$$
and at $a=1/4$ that
$$\ln\Gamma\left({1 \over 4}\right)={1 \over 2}\ln(4\pi)-{1 \over 4}+{1 \over {2\pi}}\left
\{{1 \over 2}\sum_{m=1}^\infty {{(-1)^m} \over m}[\pi-\mbox{Si}(\pi m)]+2\sum_{m=0}^\infty {{(-1)^m}\over {(2m+1)}}\mbox{Ci}[\pi(m+1/2)]\right \}.  \eqno(2.44)$$
%We omit a second direct proof of this relation.

We give a second direct proof of relation (2.43).  We have
$$\sum_{j=1}^\infty {1 \over j}[\pi-2\mbox{Si}(2\pi j)]=2\sum_{j=1}^\infty {1 \over j}
[\mbox{Si}(\infty)-\mbox{Si}(2\pi j)]=2\sum_{j=1}^\infty {1 \over j}\int_{2\pi j}^\infty
{{\sin t} \over t}dt$$
$$=2\sum_{j=1}^\infty {1 \over j}\int_0^\infty {{\sin v} \over {v+2\pi j}}dv
=2\sum_{j=1}^\infty {1 \over j}\int_0^\infty \sin v dv \int_0^\infty e^{-(v+2\pi j)x}dx$$
$$=-2\int_0^\infty \sin v dv \int_0^\infty e^{-xv} \ln(1-e^{-2\pi x})dx
=-2\int_0^\infty {{\ln(1-e^{-2\pi x})} \over {1+x^2}}dx$$
$$=4\pi\int_0^\infty {{\tan^{-1} x} \over {e^{2\pi x}-1}}dx.  \eqno(2.45)$$
In the last step we integrated by parts.  We now differentiate a known result \cite{sri}
(p. 100), to write
$$\zeta'(s)=-{1 \over {(s-1)^2}}+2\int_0^\infty {{\cos(s\tan^{-1} t)} \over {(1+t^2)^{s/2}}}
{{\tan^{-1} t} \over {(e^{2\pi t}-1)}}dt$$
$$-\int_0^\infty {{\sin(s\tan^{-1} t)} \over {(1+t^2)^{s/2}}}
{{\ln (1+t^2)} \over {(e^{2\pi t}-1)}}dt.  \eqno(2.46)$$
Since by (2.46) we have
$$\zeta'(0)=-{1 \over 2}\ln(2\pi)=-1+2\int_0^\infty {{\tan^{-1} t} \over {e^{2\pi t}-1}}dt,
\eqno(2.47)$$
the result (2.43) again follows.  

{\it Proposition 4}.  We first demonstrate part (b), with part (a) being similar.
We use the representation for Re $x>0$ (e.g., \cite{grad}, p. 928)
$$\mbox{Ci}(x) =\gamma+\ln x+\int_0^x {{(\cos t -1)} \over t}dt.  \eqno(2.48)$$
We also use the Fourier expansions
$$\sum_{n=1}^\infty {{\cos(\beta n)} \over n^4}=-{\beta^4 \over {48}}+{\pi \over {12}}
\beta^3-{\pi^2 \over {12}}\beta^2+\zeta(4),  \eqno(2.49a)$$
and
$$\sum_{n=1}^\infty (-1)^n{{\cos(\beta n)} \over n^4}=-{\beta^4 \over {48}}+{\pi^2 \over {24}}\beta^2-{7 \over 8}\zeta(4).  \eqno(2.49b)$$
Then we have
$$\sum_{n=1}^\infty {{\mbox{Ci}(\beta n)} \over n^4}=\sum_{n=1}^\infty {1 \over n^4} \left[\gamma+\ln (\beta n)+\int_0^{\beta n}{{(\cos t-1)} \over t}dt\right]$$
$$=(\gamma+\ln \beta)\zeta(4)-\zeta'(4)+\sum_{n=1}^\infty \int_0^1 {{[\cos(\beta n v)-1]}
\over v}dv$$
$$=(\gamma+\ln \beta)\zeta(4)-\zeta'(4)+{\beta^2 \over {12}}\int_0^1\left(-{\beta^2 \over 4}
v^3+\pi \beta v^2-\pi^2 v\right)dv$$
$$=(\gamma+\ln \beta)\zeta(4)-\zeta'(4)+{\beta^2 \over {12}}\left(-{\beta^2 \over {16}}+{\pi \over 3}\beta-{\pi^2 \over 2}\right).  \eqno(2.50)$$
The integral over $v$ being absolutely convergent, the interchange of summation and 
integration is justified.
%could remark on series heuristics for SIAM solution ...
For (1.22d), we use the alternating zeta function for Re $s>0$ (\cite{grad}, p. 1073, or
\cite{sri}, p. 96)
$$\sum_{n=1}^\infty {{(-1)^n} \over n^s}=(2^{1-s}-1)\zeta(s), \eqno(2.51)$$
so that
$$\sum_{n=1}^\infty {{(-1)^n} \over n^s}\ln n=(1-2^{1-s})\zeta'(s)+2^{1-s}(\ln 2)\zeta(s). \eqno(2.52)$$

For part (c) we use the summations \cite{hansen} (p. 244)  
$$\sum_{n=1}^\infty {{\cos nx} \over n^{2k}}={{(-1)^{k-1}} \over {2(2k)!}}(2\pi)^{2k}
B_{2k}\left({x \over {2\pi}}\right), \eqno(2.53a)$$
and
$$\sum_{n=1}^\infty (-1)^n{{\cos nx} \over n^{2k}}={{(-1)^{k-1}} \over {2(2k)!}}(2\pi)^{2k}
B_{2k}\left({{x+\pi} \over {2\pi}}\right). \eqno(2.53b)$$

For part (d) we use the $y=0$ case of the two summations \cite{hansen} (p. 244)
$$\sum_{k=1}^\infty {{\cos(kx+y)} \over k^a}={{(2\pi)^a} \over {2\Gamma(a)}}\csc(\pi a)\left[
\sin\left(y+{{\pi a} \over 2}\right)\zeta\left(1-a,1-{x \over {2\pi}}\right)\right.$$
$$\left.-\sin\left(y-{{\pi a} \over 2}\right)\zeta\left(1-a,{x \over {2\pi}}\right)\right], \eqno(2.54a)$$
and
$$\sum_{k=1}^\infty (-1)^k{{\cos(kx+y)} \over k^a}={{(2\pi)^a} \over {2\Gamma(a)}}\csc(\pi a)\left[\sin\left(y+{{\pi a} \over 2}\right)\zeta\left(1-a,{{\pi-x} \over {2\pi}}\right)\right.$$
$$\left. -\sin\left(y-{{\pi a} \over 2}\right)\zeta\left(1-a,{{\pi+x} \over {2\pi}}\right)\right]. \eqno(2.54b)$$

Part (e) uses 
$$\sum_{n=1}^\infty {1 \over {(2n+1)^a}}=(1-2^{-a})\zeta(a)-1, \eqno(2.55a)$$
and
$$\sum_{n=1}^\infty {{(-1)^n} \over {(2n+1)^a}}=4^{-a}\left[\zeta\left(a,{1 \over 4}\right)
-\zeta\left(a,{3 \over 4}\right)\right]-1, \eqno(2.55b)$$
together with the $y=0$ case of two summations \cite{hansen} (p. 245), so that
$$\sum_{k=0}^\infty {{\cos kx} \over {(2k+1)^a}}={{(2\pi)^a} \over {4\Gamma(a)}}\csc(\pi a)\left\{-\sin\left({{x+\pi a} \over 2}\right)\left[\zeta\left(1-a,{{x+2\pi} \over {4\pi}}\right)-\zeta\left(1-a,{x \over {4\pi}}\right)\right]\right.$$
$$\left.+\sin\left({{x-\pi a} \over 2}\right)\left[\zeta\left(1-a,{{2\pi-x} \over {4\pi}}\right)-\zeta\left(1-a,1-{x \over {4\pi}}\right)\right]\right\}, \eqno(2.56a)$$
and
$$\sum_{k=0}^\infty (-1)^k{{\cos kx} \over {(2k+1)^a}}={{(2\pi)^a} \over {4\Gamma(a)}}\csc(\pi a)\left\{\cos\left({{x+\pi a} \over 2}\right)\left[\zeta\left(1-a,{{x+\pi} \over {4\pi}}\right)-\zeta\left(1-a,{{x+3\pi} \over {4\pi}}\right)\right]\right.$$
$$\left.+\cos\left({{x-\pi a} \over 2}\right)\left[\zeta\left(1-a,{{\pi-x} \over {4\pi}}\right)-\zeta\left(1-a,{{3\pi -x} \over {4\pi}}\right)\right]\right\}. \eqno(2.56b)$$

Part (f) uses \cite{hansen} (p. 247)
$$\sum_{n=1}^\infty {{z^n \cos nx} \over {n(n+1)}}=1-{1 \over {2z}}(z-\cos x)\ln(1-2z \cos x+
z^2)-{{\sin x} \over z}\tan^{-1}\left({{z\sin x} \over {1-z\cos x}}\right).  \eqno(2.57)$$

{\it Remarks}.  In part (a), due to the functional equation of the Riemann zeta function,
we may write the value $\zeta'(2)=\zeta(2)[\gamma+\ln(2\pi)+12 \ln A]$, where $A \simeq 1.28243$ is the Glaisher constant, such that $\ln A=1/12-\zeta'(-1)=-[\zeta(-1)+\zeta'(-1)]$.

As shown by part (c), the Proposition extends to all summations $\sum_{n=1}^\infty (\pm 1)^n \mbox{Ci}(\beta n)/n^{2k}$ with $k \geq 1$. Alternatively, successive Fourier expansions may be built up by twice integrating the previous one, and requiring the appropriate $\zeta(2k)$ constant term.  Of course in this part, the relation (\cite{edwards}, p. 14 or \cite{grad},
p. 1077 or \cite{sri} p. 98) $\zeta(2k)= (-1)^{k-1}(2\pi)^{2k}B_{2k}/[2 (2k)!]$ applies. 
%and a modified form for $\sum_{n=1}^\infty(\pi/2- 
%\mbox{Si}(\beta n)/n^{2k+1}$ with $k \geq 1$.

Analogous results can be given for sums $\sum_{n=1}^\infty [\pi-2\mbox{Si}(\beta n)]/n^{2k+1}$ with $k \geq 0$.

Part (d) of the Proposition subsumes part (c), owing to the reduction $\zeta(1-n,q) =-B_n(q)/n$ (e.g., \cite{sri}, p. 85) and the functional equation $B_n(1-x)=(-1)^nB_n(x)$.  The latter relation shows the reflection property $B_{2k}(1-x)=B_{2k}(x)$.

Other Ci summations are possible, based for instance upon known sums for 
$\sum_{k=0}^\infty (\pm 1)^k a^k \cos kx/(km+n)$ \cite{hansen} (pp. 246-247).
 
%\pagebreak
\bigskip
\centerline{\bf Summary}
\medskip

We have developed a variety of series and integral representations for a family of 
functions.  From the representations for the polygamma functions follow many
special cases, including representations of the values $\psi(p/q)$, of harmonic
numbers $H_n$, and of generalized harmonic numbers $H_n^{(r)}$.  We have found 
representations for $\ln \Gamma$ and in turn a product representation for the Gamma
function.  Most of the representations are quite suited for computation, and in fact their convergence may also be accelerated.  The approach via the zeroth Stieltjes constant enables
the development of new representations, as well as the recovery of known series
and integral representations for constants such as $\gamma$ and for the digamma function.

We have given series representations in terms of trigonometric integrals Ci 
and Si for $\gamma$, for certain combinations of $\gamma$ and $\zeta(n)/n$, for $\gamma+\ln 2$, and generally for $\ln \Gamma(a)$ and $\psi(a)$.  Additionally, we have demonstrated in
several ways how to perform summations over parameterized values of Si and Ci.

%\medskip
%\centerline{\bf Acknowledgement}
%\medskip

%We thank ... for useful correspondence.  

\pagebreak

\end{document}